\documentstyle[12pt]{article}

\newcommand{\be}{\begin{equation}}
\newcommand{\ee}{\end{equation}}
\newcommand{\ba}{\begin{eqnarray}}
\newcommand{\ea}{\end{eqnarray}}

\begin{document}
\begin{flushright}
JINR preprint E2-2003-215
\end{flushright}
\begin{center}
{\bf
{Adelic Universe and Cosmological Constant}}
\end{center}
\begin{center}
 Nugzar Makhaldiani
\end{center}
\begin{center}
Laboratory of Information Technologies\\
Joint Institute for Nuclear Research\\
 Dubna, Moscow Region, Russia\\
e-mail address:~~mnv@jinr.ru
\end{center}

\vskip 5mm
\bigskip
\nopagebreak
\begin{abstract}

In the quantum adelic field (string) theory models,
vacuum energy -- cosmological constant vanish. The other (alternative ?)
mechanism is given by supersymmetric theories. Some observations on prime numbers,
zeta -- function and fine structure constant are also considered.
\end{abstract}
\newpage

{\bf 1. Introduction}\\

There is an opinion that present-day theoretical physics needs
(almost) all mathematics, and the progress of modern
mathematics is stimulated by fundamental problems of theoretical physics.

In this paper \footnote{Presented October 4 2003 at The First
International Conference on p - adic Mathematical Physics, Steclov
Mathematical Institute, Russian Academy of Sciences, Moscow,
Russia.}, I would like to show a mechanism of solving of the
cosmological constant problem \cite{Weinberg} based on the adelic
structure of the quantum field (string) theory models
\cite{Makhaldiani}. Some speculations on the fine structure
constant and the prime numbers are given.\\

{\bf 2. Cosmological constant problem }\\

The cosmological constant problem is one of the most serious paradoxes
in modern particle physics and cosmology \cite{Weinberg}.
Some astronomical observations indicate that the cosmological constant is
many orders of magnitude smaller than estimated in modern theoretical
elementary particles physics.

{\bf 2.1} In his attempt (1917), \cite{Einstein} to apply the general
relativity to the
whole universe, A.~Einstein invented a new term involving a free parameter
$\lambda$, the cosmological constant (CC), \ba
R_{\mu\nu}-\frac{1}{2}g_{\mu\nu}=\lambda g_{\mu\nu}-8\pi
GT_{\mu\nu}. \ea

With this modification he finds a static solution
for the universe filled with dust of zero pressure and mass density
\ba \rho=\frac{\lambda}{8\pi G}. \ea

The geometry of the universe was that of a sphere $S_3$ with proper
circumference $2\pi r,$ where \ba r=\lambda^{-1/2}, \ea so the
mass of the universe was \ba
&M=2\pi^2r^3\rho=\frac{\pi}{4}G^{-1}\lambda^{-1/2}\cr &\sim r
(?!). \ea

Any contributions to the energy density of the
vacuum acts just like CC. By Lorentz invariance, in the vacuum,
\ba <T_{\mu\nu}>=-<\rho>g_{\mu\nu}, \ea so \ba
\lambda_{eff}=\lambda+8\pi G<\rho>,
\ea
or the total vacuum energy density
\ba
\rho_V=<\rho>+\frac{\lambda}{8\pi G}=\frac{\lambda_{eff}}{8\pi G}.
\ea
The experimental upper bound on $\lambda_{eff}$ or $\rho_V$ is
provided by measurements of cosmological redshifts as a function
of distance. From the present expansion rate of the universes
\cite{PD} \ba \frac{dlnR}{dt}\equiv H_0=100h\frac{km}{secMpc},\ \
h=0.7\pm 0.07 \ea we have \ba H_0^{-1}=(1\div 2)\times 10^{10}ye,\
\
 |\lambda_{eff}|\leq H_0^2,\ \   |\rho_V|\leq 10^{-29}g/cm^2\simeq10^{-47}GeV^4.
\ea

{\bf 2.2} The quantum oscillator with hamiltonian \ba
H=\frac{1}{2}P^2+\frac{1}{2}\omega^2x^2, \ea
has the energy spectrum
\ba E_n=\hbar \omega(n+1/2), \ea with the lowest, vacuum,
value $E_0=\hbar \omega$. Normal modes of a quantum field of mass
m are oscillators with frequencies $\omega (k)=\sqrt{k^2+m^2}.$
Summing the zero-point energies of all normal modes of the field
up to a wave number cut-off $\Lambda>>m$ yields a vacuum energy
density \ba <\rho>=\int\limits_{0}^{\Lambda}\frac{4\pi
k^2dk}{(2\pi)^3} \frac{1}{2}\sqrt{k^2+m^2}\simeq
\frac{\Lambda^4}{16\pi^2}. \ea
If we take $\Lambda=(8\pi
G)^{-1/2},$ then
\ba <\rho>\simeq2^{-10}\pi^{-4}G^{-2}=2\times
10^{71}GeV^4. \ea We saw that \ba |<\rho>+\frac{\lambda}{8\pi
G}|\leq10^{-47}GeV^4\simeq(10^{-3}eV)^4, \ea so the two terms must
cancel to better than 100 decimal places! If we take
$\Lambda_{QCD}$, $<\rho>\simeq 10^{-6}GeV^4$, the two terms must
cancel better to than 40 decimal places. Since the cosmological
upper bound on $<\rho_{eff}>$ is vastly less than any value
expected from particle theory, theorists assumed that (for some
unknown reason) this quantity is zero.\\

{\bf 3. Supersymmetric mechanism of solution to the CC problem }\\

A minimal realization of the algebra of supersymmetry
\ba \label{QH}
&&\{Q,Q^+\}=H,\cr
&&\{Q,Q\}=\{Q^+,Q^+\}=0,
\ea
is given by a point particle in one dimension, \cite{Witten}
\ba \label{QQ+}
&& Q=a(-iP+W),\cr
&&Q^+=a^+(iP+W),
\ea
where $P=-i\partial /\partial x$, the superpotential $W(x)$ is any function
of x, and spinor operators $a$ and $a^+$ obey the anticommuting relations
\ba \label{phi}
&& \{a,a^+\}=1, \cr
&& a^2=(a^+)^2=0.
\ea

There is a following representation of operators
$a$, $a^+$ and $\sigma$ by
the Pauli spin matrices
\ba \label{Pauli}
a&=&\frac{\sigma_1-i\sigma_2}{2}, \cr
a^+ &=&\frac{\sigma_1+i\sigma_2}{2}, \cr
\sigma&=&\sigma_3.
\ea

From formulae (\ref{QH}) and (\ref{QQ+}) then we have
\ba \label{H}
H=P^2+W^2+\sigma W_{x}.
\ea

The simplest nontrivial case of the superpotential $W=\omega x$
corresponds to the supersimmetric oscillator with Hamiltonian \ba
H=H_B+H_F, \ \ H_B=P^2+\omega^2x^2,\ \ H_F=\omega\sigma, \ea wave
function \ba \psi=\psi_B\psi_F, \ea and spectrum \ba
H_B\psi_{Bn}=\omega(2n+1)\psi_{Bn},\cr H_F\psi_+=\omega\psi_+,\ \
H_F\psi_-=-\omega\psi_-. \ea
The ground state energies of the bosonic
and fermionic parts are \ba E_{B0}=\omega,\ \ E_{F0}=-\omega, \ea
so the vacuum energy of the supersymmetric oscillator is
\ba
<0|H|0>=E_0=E_{B0}+E_{F0}=0,\ \ |0>=\psi_{B0}\psi_{F0}.
\ea

{\bf 3.1} Let us see on this toy - solution of the CC problem
from the quantum
statistical viewpoint. The statistical sum of the
supersymmetric oscillator is \ba Z(\beta)=Z_BZ_F, \ea where \ba
Z_B=\sum_{n}^{}e^{-\beta E_{Bn}}=e^{-\beta
\omega}+e^{-\beta\omega(2+1)}+...\cr Z_F=\sum_{n}^{}e^{-\beta
E_{Fn}}=e^{\beta \omega}+e^{-\beta\omega}. \cr \ea In the low
temperature limit, \ba Z(\beta)=1+O(e^{-\beta 2\omega})\rightarrow
1,\ \ \beta =T^{-1}, \ea so CC \ba \lambda\sim
lnZ\rightarrow 0. \ea
{\bf 3.2} In the case of the adelic solution to the CC
problem we will have, \ba
Z(\beta)=\prod_{p\geq1}^{}Z_p=Z_1Z_2Z_3Z_5...,\cr Z_1\equiv Z_B,\
\ Z_F\div Z_2Z_3Z_5... (?!) \ea \\

{\bf 4. p - adic fractal calculus
and adelic solution of the cosmological constant problem}\\

Every (good) school boy/girl knows what is \ba \frac{d^n}{dx^n},
\ea but what is its following extension \ba
\frac{d^{\alpha}}{dx^{\alpha}}= \ ?,\ \ \alpha\in R. \ea

Let us
consider the integer derivatives of the monomials \ba
\frac{d^n}{dx^n}x^m&=&m(m-1)...(m-(n-1))x^{m-n},\ \ n\leq m,\cr
           &=&\frac{\Gamma(m+1)}{\Gamma(m+1-n)}x^{m-n}.
\ea
L.Euler (1707 - 1783) invented the following definition
of the fractal derivatives:
\ba
\frac{d^{\alpha}}{dx^{\alpha}}x^{\beta}=
\frac{\Gamma(\beta+1)}{\Gamma(\beta+1-\alpha)}x^{\beta-\alpha}.
\ea
J.Liuville (1809-1882) takes exponentials as a base functions,
\ba
\frac{d^{\alpha}}{dx^{\alpha}}e^{ax}=a^{\alpha}e^{ax}.
\ea
J.H.~Holmgren (1863) invented the following integral transformation
\ba \label{Holmgren}
D_{c,x}^{-\alpha}f=\frac{1}{\Gamma(\alpha)}
\int\limits_{c}^{x}|x-t|^{\alpha-1}f(t)dt.
\ea

It is easy to show that \ba
D_{c,x}^{-\alpha}x^m&=&\frac{\Gamma(m+1)}{\Gamma(m+1+\alpha)}
(x^{m+\alpha}-c^{m+\alpha}),\cr
D_{c,x}^{-\alpha}e^{ax}&=&a^{-\alpha}(e^{ax}-e^{ac}), \ea
so
$c=0,$ when $m+\alpha \geq 0,$ in Holmgren's definition of the
fractal calculus, corresponds to the Euler's definition, and
$c=-\infty$, when $a>0,$ corresponds to the Liuville's definition.

Note also the following slight modification of the $c=0$ case \cite{Mfractal}
\ba \label{Mfractal}
&&D_{0,x}^{-\alpha}f=\frac{|x|^{\alpha}}{\Gamma(\alpha)}
\int\limits_{0}^{1}|1-t|^{\alpha-1}f(xt)dt \cr
&&=\frac{|x|^{\alpha}}{\Gamma(\alpha)}B(\alpha,\ \frac{d}{dx}x)f(x)=
|x|^{\alpha}\frac{\Gamma(\frac{d}{dx}x)}{\Gamma(\alpha+\frac{d}{dx}x)}f(x),\cr
&&f(xt)=x^{t\frac{d}{dt}}f(t)=t^{x\frac{d}{dx}}f(x), (\frac{d}{dx}x)^{-1}=
x^{-1}\int\limits_{0}^{x}dx.
\ea
Let us consider integer derivatives, $\alpha=-n,$
\ba
&&D_{0x}^nf=\frac{1}{x^n}\frac{\Gamma(\partial x)}{\Gamma(-n+\partial x)}f\cr
&&=x^{-n}(-n+\partial x)(-n+1+\partial x)...(-1+\partial x)f=...\cr
&&=x^{-n}(-n+1+x\partial)x^{n-1}{\partial }^{n-1}f=f^{(n)}.
\ea
Fractal derivatives, when $\alpha=-n+\varepsilon, 0<\varepsilon<1,$ can
be calculated as $D^nD^{-\varepsilon}.$\\
It is easy to show, that $D^{-k-1}f=D^{-k}D^{-1}f,$ so integrals can be
calculated as $D^{-n}f=(D^{-1})^nf,$ where
\ba
D^{-1}f=x\frac{\Gamma(\partial x)}{\Gamma(1+\partial x)}f=x\frac{1}{\partial x}f=
(\partial)^{-1}f.
\ea

{\bf 4.1} As an example, let us consider Weierstrass C.T.W. (1815 - 1897)
fractal function \ba f(t)=\sum_{n\geq
0}^{}a^ne^{i(b^nt+\varphi_n)},\ \ a<1,\ \ ab>1. \ea

For fractals
we have no integer derivatives, \ba
f^{(1)}(t)=i\sum_{}^{}(ab)^ne^{i(b^nt+\varphi_n)}=\infty, \ea
but
the fractal derivative,
\ba
f^{(\alpha)}(t)=i^{\alpha}\sum_{}^{}(ab^{\alpha})^ne^{i(b^nt+\varphi_n)},
\ea when $ab^{\alpha}=a'<1,$ is another fractal \cite{Mfractal}.

{\bf 4.2} Definition of the p-adic norm, $|\  |_p$ for raitional
numbers $r\in Q$ is
\ba
&|r|_p=p^{-k},\  r\neq 0;\cr
&|0|_p=0,
\ea
where $k=ord_p(r)$ is defined from the following representation of the r
\ba
r=\pm p^{k}\frac{m}{n},
\ea
integers m and n do not contain as factor p.

p-adic analog of the fractal calculus (\ref{Holmgren}) ,
\ba\label{Vladimirov}
D_x^{-\alpha}f=\frac{1}{\Gamma_p(\alpha)}\int\limits_{Q_p}^{}|x-t|_p^{\alpha-1}
f(t)dt,
\ea
where $f(x)$ is a complex function of the p-adic variable x,
with p-adic $\Gamma$--function
\ba
\Gamma_p(\alpha)=\int\limits_{Q_p}^{}dt|t|_p^{\alpha-1}\chi (t)=
\frac{1-p^{\alpha-1}}{1-p^{-\alpha}},
\ea
was considered by V.S.~Vladimirov \cite{Vladimirov}.

Note also the following slight modification of (\ref{Vladimirov}),
\ba\label{pfractal}
D_x^{-\alpha}f=\frac{|x|^{\alpha}_p}{\Gamma_p(\alpha)}
\int\limits_{Q_p}^{}|1-t|_p^{\alpha-1}f(xt)dt=
|x|^{\alpha}_p\frac{\Gamma_p(\partial |x|)}{\Gamma_p(\alpha+\partial |x|)}f(x).
\ea
Last expression is applicable for functions of type $f(x)=f(|x|).$

{\bf 4.3} Let us consider the following action
\ba\label{Sp}
S=\frac{1}{2}\int\limits_{Q_v}^{}dx \Phi (x)D_x^{\alpha}\Phi,\ v=1,2,3,5,... \ea

In the momentum representation
\ba
S=\frac{1}{2}\int\limits_{Q_v}^{}du\tilde{\Phi}(-u)|u|_v^{\alpha}\tilde{\Phi}(u),
\ea
where
\ba
\Phi(x)=\int\limits_{Q_v}^{}du\chi_v(ux)\tilde{\Phi}(u),\cr
D^{-\alpha}\chi_v(ux)=|u|_v^{-\alpha}\chi_v(ux).
\ea

The statistical sum of the corresponding quantum theory is
\ba
Z_v=\int\limits_{}^{}d\Phi e^{-\frac{1}{2}\int\limits_{}^{}\Phi D^{\alpha}\Phi}
=det^{-1/2}D^{\alpha}=(\prod_{u}^{}|u|_v)^{-\alpha/2}.
\ea

Note that, by fractal calculus and vector generalization of
the model (\ref{Sp}),
string amplitudes were obtained in \cite{Makhaldiani}.

{\bf 4.4} Adels $a\in A$ are constructed by real $a_1\in Q_1$ and
p-adic $a_p\in Q_p$ numbers (see e.g. \cite{Gelfand})
\ba
a=(a_1, a_2, a_3, a_5, ..., a_p, ...),
\ea
with restriction that $a_p\in Z_p=\{ x\in Q_p, |x|_p\leq 1 \}$ for all
but a finite set F of primes p.

$A$ is a ring with respect to the componentwise addition and multiplication. A
prinsipal adel is a sequence $r=(r, r, ..., r, ...)$, $r\in Q$-rational number.

Norm on adels is defined as
\ba
|a|=\prod_{p\geq 1}^{}|a_p|_p.
\ea
Note that the norm on principal adels is trivial.

In the adelic generalization of the model (\ref{Sp}),
\ba
\Phi(x)=\prod_{p\geq 1}^{}\Phi_p(x_p),\ \ dx=\prod_{p\geq 1}^{}dx_p,\ \ D_x^{\alpha}
=\sum_{p\geq 1}^{}D_{x_p}^{\alpha},
\ea
where by $D^{\alpha}_{x_1}$ we denote fractal derivative (\ref{Mfractal}),
$x_1$ is real and
$|\ |_1$ is real norm.
If
\ba
\int\limits_{}^{}dx_p|\Phi(x_p)|^2=1,
\ea
then
\ba
\int\limits_{}^{}dx|\Phi(x)|^2=1,\ \ S=\sum_{p\geq 1}^{}S_p,
\ea
so
\ba
Z=\prod_{p\geq 1}^{}Z_p=\prod_{p\geq 1}^{}(\prod_{u}^{}|u|_p)^{-\alpha/2}=
(\prod_{u}^{}\prod_{p\geq 1}^{}|u|_p)^{-\alpha/2}=1,\ \
\lambda\sim lnZ=0,
\ea
if $u\in Q$.\\

{\bf 5. Some observations on zeta function, prime numbers and fine
structure constant} \\

Extended particles: nuclei, hadrons, strings,...
are characterized by exponential state density
\ba
\rho(E)\sim e^{\beta_H E}.
\ea
Gas of the extended particles described by statistical sum
\ba
Z=\sum_{n}^{}e^{-\beta E_n}=\sum_{E_n}^{}\rho (E_n)e^{-\beta E_n},
\ea
is well defined for $\beta\geq\beta_H$ or $T\leq T_H=1/\beta_H$ - Hagedorn
temperature (see e.g. \cite{Makh2}).

{\bf 5.1} The following representations of zeta-function \cite{Titchmarsh}
\ba \zeta(\beta)=\sum_{n\geq
1}^{}\frac{1}{n^{\beta}}=\sum_{n\geq 1}^{}e^{-\beta E_n}
=\prod_{p\geq 2}^{}\frac{1}{1-p^{-\beta}}=\prod_{p\geq 2}^{}\zeta_p, \ea where
$E_n=lnn$, are defined for Re$\beta>1$.

In physical terms,
zeta-function is almost a statistical sum of ideal gas of quantum
bosonic oscillators with frequencies $\omega =lnp$. The following
modification of the partial zeta-functions, \ba\label{B1}
Z_{pB}=p^{-\beta/2}\zeta_p(\beta)=\frac{p^{-\beta}/2}{1-p^{-\beta}}
=\frac{1}{p^{\beta/2}-p^{-\beta/2}}, \ea corresponds exactly to
the quantum bosonic oscillators.

Zeta-function has a pole at $\beta=1$, "trivial" zeros at
$\beta=-2n, n\geq1$ and, according to Riemann's hypothesis,
nontrivial (complex) zeros on the imaginary line
$\beta=1/2+i\lambda_n.$

{\bf 5.2} In a sense  the following
reciprocal zeta-function looks  more interesting (less reducible):
\ba\label{R}
\zeta_r(\beta)=\frac{1}{\zeta(\beta)}=\prod_{p}^{}(1-p^{-\beta})
=\sum_{n\geq 1}^{} \frac{\mu(n)}{n^{\beta}}=(1-\beta)R(\beta). \ea
Hhere $\mu(n)$-Mobius arithmetic function is defined on
natural numbers as
\ba \mu(1)=1, \ \ \mu(n)=(-1)^k, \ea
if the
factorized form of n, $n=p_1p_2...p_k$ contains only different prime
factors and is zero if two factors coincide. Partial reciprocal
zeta-functions, \ba\label{F}
\zeta_{pr}(\beta)=1-p^{-\beta}=e^{-\beta\omega/2}Z_{pF}(\beta),
\ea almost coincide with the fermionic oscillator statistical
sum,
\ba\label{F1}
Z_{pF}(\beta)=\sum_{E_n}^{}\rho(E_n) e^{-\beta
E_n}=\sum_{E_n}^{}e^{-\beta F_n}, \ea
where the density of the
occupied fermionic state is negative \ba \rho(E_1)=-1, \ea free
energy $F_n$ and entropy $S_n$ are \ba F_n=E_n+S_nT, \ \
E_n=\omega(n-1/2), \ \ S_n=i\pi n,\ \ \omega=lnp,\ \ n=0;1. \ea

We can
consider mixed quantum gases with different primes if we restrict
ourselves with some maximal prime $p_N$, \ba
Z_{NB}=\prod_{p=p_1}^{p_N}Z_{pB},\ \ Z_{NF}=\prod_{p=p_1}^{p_N}Z_{pF},
\ea
but we cannot consider the quantum systems with the infinite number
of prime components without renormalization (simply neglecting)
infinite vacuum energy.

For $\zeta_r$-functions we have an adelic identity
\ba \prod_{p\geq 1}^{}\zeta_{pr}=1,\ \ \zeta_{1r}\equiv \zeta, \ea
so in the
corresponding, "number - theoretic universe"there is not
a CC-problem.

Note that the quantum statistical sums (\ref{B1},\ref{F1}) are antisymmetric
with respect to the dual transformation $p\rightarrow p^{-1}.$
Physical quantities, which are logarithmic derivatives of the
statistical sums, remain invariant. The classical limit,
$p\rightarrow 1,$ corresponds to the selfdual point p=1.

{\bf 5.3} Following extension of the integer numbers

\ba
[n]_p=\frac{p^n-1}{p-1}=1+p+p^2+...+p^{n-1},
\ea
represents repunits (see e.g. \cite{[n]}),
\ba
[n]_p=11...1.
\ea
In the classical limit, $p\rightarrow 1,$  $[n]_1=n.$
Note also the identity
\ba
[p_1p_2....p_k]_q=[p_1]_q[p_2]_{q^{p_1}}...[p_k]_{q^{p_1p_2...p_{k-1}}}.
\ea
This identity in the classical limit,$q\rightarrow 1$, reduce to the
main arithmetic relation $n=p_1p_2...p_k.$
If we take $q=exp(\frac{2\pi i}{p}),$ then $[n]_q=0,$ when $p$ is equal to
one of the factors of $n$.

{\bf 5.4} Now, for a hadronic
string model (see e.g. \cite{Barbashov}) we know,
that the high temperature phase, $T>T_H,$ is the
quark-gluon phase or, as it was named by S.B.~Gerasimov, Gluqua.

Interesting questions are:\\
$\bullet$ what is the high temperature phase of the
fundamental string (Twistor; Topological; p-adic...) ?\\
$\bullet$ What is the "high temperature phase", $\beta\leq 1,$ of
the zeta-function, what are the constituents of the (prime)
numbers ?

The following identity
\ba
\frac{1}{1-x}=(1+x)(1+x^2)(1+x^4)...
\ea
 for $x=p^{-\beta}$ tells us that (almost) bosonic gas
of prime oscillators can be represented as a gas of (almost)
fermionic oscillators with frequencies $\omega=2^nlnp.$ This is a
hint on a grassmann constituents
of primes.\\

{\bf 5.5} Function $R(\beta)$ defined in (\ref{R}) has
the poles in the same points where zeta-function has zeros. So it
is natural to investigate R-function by methods of scattering
theory \cite{SCAT}. Corresponding resolvent \ba \hat
R(\beta)=\frac{1}{\beta-\hat H}, \ea
defines a hamiltonian with eigenvalues as zeros of zeta-function.

{\bf 5.6} For each prime p we have the following representation of
$-1$ \ba \label{-1} -1=(p-1)(1+p+p^2+p^4+...),\ea so we can
eliminate negative numbers in the field of p-adic numbers, for
each p. Now we can represent $\sqrt{-1}$
\ba
i=\sqrt{-1}=\sqrt{p-1}\sqrt{1+p+...}.\ea

Thus, for some primes, \ba
p=4k^2+1=5,\ \ 17,\ \ 37,\ \ 101,\ \ 197,\ \ 257,\ ...\ea
we can also eliminate complex numbers. Next,
$\sqrt[4]{-1}$ can be eliminated for primes \ba
p=2^{2^2}k^{2^2}=17,\ \ 257,\ ...\ea and $\sqrt[8]{-1}$ can be
eliminated for primes  \ba p=2^{2^3}k^{2^3}+1=257,\ \ ...\ea

Note
that the nearest integer to prime 257 is $256=2^8=1$byte.

 Let me also
mention that in quantum computing (Quanputing, \cite{Makh3}) we already have
quantum logic (dynamics, algorithms,...) but have not yet quantum
ethic (save conditions for quanputation, decoherence problems).

In a more general case, $\sqrt[2^n]{-1},$ we come to the primes
\ba\label{Ferma}
p=2^{2^n}k^{2^n}+1. \ea
The case $k=1$ in (\ref{Ferma}) corresponds to the primes
of Fermat(1601 - 1665).

{\bf 5.7} In quantum electrodynamics \cite{BSh}, there is a fundamental
constant $\alpha$-fine structure constant. The value of
$\alpha^{-1}=137.036...$ \cite{PD} is in a good approximation given by prime
p=137\footnote{Another prime number that I like is 887 - lifetime
of neutron in seconds \cite{PD}. I like that $137+887=1024=2^{10}=1K.$ }.
There is no theoretical explanation to this value.

Note that \ba 137=11^2+4^2=|11+4i|^2=|4+11i|... \ea

Now a curious
question is: what is the distance between $z_1=11+4i$ and
$z_2=4+11i,$ \ba &&|z|=|z_1-z_2|=\sqrt{49+49}=\sqrt{100-2}=
10(1-\frac{1}{2}\frac{2}{100}+...).\\
&&|z|=10-O(1\%)
\ea

 If we want to take exactly 10, we must rise 11 a little.
This will be in right direction, but gives for
$\alpha^{-1}=138.5...$ So for more precise value of $\alpha^{-1}=137.036...,$
we will have a
little bigger value of $|z|$, but less than 10. Let us see also a toy-version
of the previous consideration,
\ba
&&37=6^2+1=|6+i|=|1+6i|=...\cr
&&|z|=\sqrt{5^2+5^2}=\sqrt{49+1}=7+O(1\%),\cr
&&37+87=124.
\ea

If we put on the complex plane all the eight points $z_1,
z_2,...,z_8,$ and connect the nearest points, we obtain an eightangle
with sites with lengths 8 and (almost)10 (2 and 7, in the toy model).
It seems interesting
that with this figure we can cover the plane on the scale of 10
figures, then the deviation of order 1 (fundamental) unit of length
appears. Next characteristic scale is of an order of 100 figures,
where deviation of the scale of 1 figure appears\footnote{There are some
other similar structures, e.g. $12^2+12^2=17^2-1.$ This way we come to the
rational approximations of the $\sqrt 2=
1.41421...$}.

Some
characteristic scales of the quantum theory of particles are :
atomic scale $\sim 10^{-8}cm,$ quantum electrodynamic scale $\sim
10^{-11}cm,$ strong interaction scale $\sim 10^{-13}cm,$ week
interaction scale $\sim 10^{-16}cm,$ Plank scale $\sim 10^{-33}cm.$
There are also other scales including macroscopic and cosmological
scales.\\

{\bf 5.8} Dirac-Schwinger's quantization \cite{D,S}
\ba\label{DS} eg=n, \ea
says that if there is in Nature even one magnetic monopole, with charge g,
electric charge e is quantized. From (\ref{DS}), when n=1, we see
\ba
\alpha_{g}=g^2=e^{-2}=\alpha^{-1}=137, \ea and fundamental
force between elementary monopoles is \ba
F_g=\frac{g^2}{r^2}=\frac{137}{r^2}.\ea

{\bf 6. Conclusions and perspectives} \\

There were different attempts to solve the CC-problem (see e.g.\cite{Weinberg}),
one of them is on the way of introduction of the several time
coordinates \cite{Dragovich}.

The adelic mechanism considered in this paper can be included also
in the adelic generalization of the standard model of cosmology
\cite{Dragovich Volovich}.

Zeta-function considerations
in this text contain a hint that there is a modification of the
quantum field theory not containing divergences.

Now let me draw a general picture and make some hand-waving arguments.
Classical fields (see e.g. \cite{Rubakov}) have a
phenomenological, macroscopic, meaning. If we want to understand their
fundamental, microscopic, structure, we might find their quants, quantize
the fields (see e.g. \cite{BSh}). If we have problems with quantization of
a field,
(could not construct corresponding perturbation theory; strong coupling;
unrenormalizable interactions, ...) it maybe due to nonfundamental,
composite nature of the field,
and we should try to find corresponding constituent fields. Sometimes we
are forced to take nonconvenient start: with free field (string) theory
tachions (Higgs particles ! Neutrinos ?)  (fundamental
bosonic string ground state); Big Bang...

In the functional integral formulation \cite{GJ} of the quantum
(as well as classical) theory and renormgroup method \cite{BSh,W}, for
consistency, we should have ultraviolet and infrared attractors (usualy
fixed points, but they maybe as well quasi-periodic cycles and even strange
attractor-fractals) in the space of coupling constants and fields.

CC--problem is cosmological problem, so we expect, that due to the
renormgroup evolution, on the cosmplogical scale, we have an
effective theory with supersymmetry and/or adelic structure.

I would like to thank V.S.~Vladimirov, I.Ya.~Aref'eva,
I.V.~Volovich and B.~Dragovich for helpful discussions on
the subject of this paper.


\end{document}